# Self-Repairing Energy Materials:
# *Sine Qua Non* for a Sustainable Future

David CAHEN* and Igor LUBOMIRSKY*

Dept. of Materials & Interfaces, Weizmann Institute of Science, Rehovot, Israel 76100

Materials are central to our way of life and future. Energy and materials as resources are connected and the obvious connections between them are the energy cost of materials and the materials cost of energy. For both of these resilience of the materials is critical; thus a major goal of future chemistry should be to find *materials for energy that can last longer*, i.e., design principles for self-repair in these.

- david.cahen@weizmann.ac.il          Orcid# 0000-0001-8118-5446
- igor.lubomirsky@weizmann.ac.il       Orcid# 0000-0002-2359-2059

Materials are central to our way of life and to its and our future. Earlier we noted [1] how closely energy and materials as resources are connected, concluding essentially that even if we have unlimited supply of one, that does not guarantee such supply of the other. The obvious connections between the two are the energy cost of materials and the materials cost of energy. For both of these resilience of the materials is a critical element. While possibly not to the liking of proponents of a consumer society we view *materials that provide the function for which they are chosen and used, longer,* as a major goal of modern chemistry and a necessary condition ("*sine qua non*") of a sustainable society.

To that end we need to know the design principles for self-repair in materials, systems and device structures. Some such principles are known and in most of today's self-repairing (self-healing) material/ systems this is achieved by [2]

1. macroscopic separation of affected part from the rest that remained functional and / or
2. replacing the affected part by a new functional part, using (finite) stored means.

Ideally, though, one wants no (or minimal) use of free energy for self-repair. This is very different from repair in living systems, often viewed as a model for self-repair and resilience of materials [2], where the availability of free energy (from food) is what allows the system to make the repairs.

What is very much common with living systems is the focus on maintaining /returning the affected part to (close to) its original functional level.

Beyond the isolated material there is the system and here electronics provide an example of another principle, namely redundancy, which, also can be viewed as biomimetic. If the level of redundancy in an electrical circuit (or, as in the early Tesla car, of a system of batteries) is sufficient then only -1- above can suffice to correct for functional errors.

A problem here is that as the level of complexity increases, so does the cost of including repair mechanisms such as -1- or -2- and at some point that cost will become prohibitive, money and energy-wise, further removing us from a *sustainable* manufacturing-use-recycle reality.

*Sustainability* requires non-destructive use of resources, which is not possible from the entropy point of view. However, since solar and geothermal energy can be viewed as "infinite" resources, what is needed for sustainability is wise use of what we call renewable energies, which then serves to bypass the entropy trap.

*Sustainability* furthermore requires materials for using the available energy sources. [1]As the amount of materials on earth is limited and 100% recycling is impossible, we need to use earth-abundant materials as well and as efficiently as possible. Such goal requires not only improving recycling but also minimizing materials consumption by cutting down waste in fabrication (e.g., via bottom-up processes such as 3-D printing) and extending their operational lifetime.[3] The latter can be achieved by self-repair of materials and material systems for energy conversion and storage. Energy conversion materials include catalytic materials for batteries, fuel cells and (solar) fuel syntheses, materials for lighting, and for photovoltaics. Naturally, energy conversion systems also contain materials for support, mechanical strength, isolation, and for guiding the heat, electricity, light or chemicals. For instance, development of underhydrated cement, capable of "healing" the surface cracks due to reaction with the atmospheric, increases its lifetime several-fold (ch. 5 in [2]-c).

As noted already self-healing is central to living systems and their growth process; it requires free energy to keep the system out of equilibrium. In inanimate systems self-repair refers to restoring original functions that were harmed. A general way to do so is -1- above, exclusion of the damaged part to allow continuing operation of the rest of the system (and some DNA repair functions this way, involving proteins and chemical energy input [4]). This approach is well known on both macroscopic and microscopic levels. On the macroscopic level damaged units are switched off without affecting overall system functions, a principle used in electrical systems. On the microscopic level it is realized when a system changes to prevent further degradation. For instance, damage to Si solar cells in an oxygen-containing environment will be limited by natural oxide growth that isolates and limits damaged areas, preventing failure of the whole panel, a process that likely helped Hubble's solar panels maintain output notwithstanding many meteorite hits (cf. Fig. 1). In some organic solar cells and LEDs this is achieved by introducing a special polymer that is activated by high currents and blocks the areas where short-circuiting occurs.[5]

In organic polymers, self-repair can be achieved by incorporating capsules with a "healing agent".[2] Similar effects are observed in self-healing of mechanically damaged GaAs nanowires in vacuum, i.e., local electrostatic effects cause surfaces to reconstruct into the original lattice. As such these are examples of non-autonomous self-repair, requiring an external influence, and in contrast to autonomous self-repair, [2] -a which we will discuss below.

Often the primary degradation mechanism of inorganic, non-molecular materials (mechanical, electrical, chemo-mechanical and radiation fatigue) is that of generation of point defects at the surfaces and in the bulk of *structurally ordered* materials, followed by coalescence of the point defects into extended ones. Because generation of point defects is unavoidable above 0 K, promoting recombination of these defects to reconstitute the original material is one possible route to extend the lifetime of materials, especially those used as active components for large-scale energy conversion. Defect recombination may occur by thermally activated processes, with defect migration or non-activated ones (like "blinking vacancies"). In general, monitoring "transient" processes of this kind is complicated, but recent advances in experimental techniques may become part of the toolkit to identify, understand, develop and tune self-repair mechanisms in materials and systems, relevant to energy. In the following, we give some examples that are already known, not only to illustrate that the goal is not an impossible one, but also because they serve to provide an idea of possible roads that chemists can explore to approach this holy grail.

*Minimize materials fatigue by introducing high densities of mobile point defects:* Operation of many energy-conversion devices, electrodes for batteries and fuel cells, micro-electromechanical energy harvesters (piezoelectric, magnetic) and switches, is associated with development of large strains (0.1-3%). In electrodes for batteries and fuel cells, the strain is chemo-mechanical in nature. In the energy harvesters and switches, the strain is mechanical or electro-mechanical.

In all cases, the strain causes material fatigue and subsequent failure via condensation of point- into line defects into cracks and voids. An accepted way to mitigate the fatigue is by nano-structuring the materials or restricting the strain, which sacrifices device performance to improve reliability. Developing self-healing materials will offer an alternative. One might thus want to target materials with significant mobile point defect densities, i.e., ionic/atomic species that can move in the material at the device's operating temperature. Such materials usually exhibit large inelastic effects that prevent strain concentration, impeding crack formation and propagation. As a result, such materials may withstand unusually large strains for many cycles. This approach is especially promising for electrode materials and solid electrolytes, where high ionic conductivity and, therefore, high point defect density, is required anyway.

In practice, we can use a recently proposed principle of "stress-adaptation", [6] based on introducing dopants that, while not hurting the materials' performance, allow it to change a linear dimension anelastically either via association/dissociation of point defects or by reorienting elastic dipoles created by these effects. For instance, self-supported membranes films of Gd-doped ceria show remarkable resistance to mechanical fatigue, supporting this idea and demonstrating the feasibility of miniature fuel cells (Fig. 2). Correlating the defect densities, defect mobility and the ability of the material to resist fatigue may well lead to materials with suppressed fatigue because of self-repair.

*Photovoltaic cells* absorb photons of energy higher than their optical bandgap. Because for most (but not all!) semiconductors the valence band is identified with bonding orbitals and the conduction band with antibonding ones, such photoexcitation across the band gap, which excites an electron from the valence into the conduction band, mostly is a process that weakens chemical bonds in the material. For all semiconductors such photon absorption generates significant concentrations of strong oxidizing and reducing agents (electrons and holes), explaining some of the destabilizing influences to which materials in solar cells are subjected.[7] Furthermore small mechanical changes may occur due to day-night temperature differences and differences in thermal expansion. In materials *with significant concentrations of point defects that are mobile during operating temperatures*, these defects can repair damage by re-establishing the atomic equilibrium that was disturbed by light absorption. The natural day/night cycle allows also relatively slow (hrs) processes to heal damage. An example of such a material is $Cu(In,Ga)Se_2$, CIGS, where we explained experimental results (on the parent compound, $CuInSe_2$) by self-healing of light and particle beam-induced defects, based on our finding that the Cu ion has some near room temperature mobility leading to relatively fast recombination of the defects.[8] (Fig. 3).

*Lattice dynamics as means for self-repair of defects.* A possibly related mechanism is to be found in the $AMX_3$ family of halide (X) perovskites, on which quite efficient PV cells and LEDs are based. We speculated on the presence of self-repair during preparation, because of their remarkably low (electrically and optically active) defect density, even though they can be prepared at relatively low temperatures, below 150 C.[9] A possible source for such process is, in this case, in addition to the static disorder that accompanies lattice defects, *dynamic* disorder, [10] which can allow dynamic adaptive self-repair. The highly anharmonic motions of atoms impart flexibility (i.e., tendency to spontaneously, make and break bonds) and thus, eject (to the surface) or absorb defects upon their formation. Recent vibrational spectroscopy results provide support for this hypothesis. One can then search for other material types with heavy atom backbones (e.g., $Cs_2HgI_4$ with $Sr_2GeS_4$ structure and Cs Lanthanide halides), to explore this design guide. These examples illustrate roads to the use of what at first glance may seem non-ideal materials for energy conversion.

In <u>*light-emitting electrochemical cells (LECs)*</u> the high (well above bandgap) voltages, required to operate light-emitting diodes at sizeable currents provide ample energy for damage, which puts stringent requirements on the materials that can be used.[11]  In LECs, the system (hitherto doped polymers) only becomes a diode under applied voltage and, once this voltage is removed, returns to its (equilibrium) *status quo ante*, allowing repair of damage. This principle, of healing damage by diffusion, maybe viewed as another avenue for self-repair in materials designed for strong non-equilibrium functions.

Extending the concept of self-repair beyond the polymer-, ceramic and metallic materials that are known and studied until now,[2] and focusing on autonomous self-repair that is, in a way, **anti-biomimetic**, will go against some presently accepted notions in this area, but holds the promise of significantly improving our chance to make durable and long-lived sustainable energy systems possible. Such effort will need to be complemented by use of more "classical" self-repairing systems, such as use of plastics and concrete of compositions that can (non-autonomously) heal cracks. [2]

Given the central role of materials in determining the feasibility of new energy supply options, and the clearly finite supply of many of the elements, we argue that it is imperative to complement recycling with longer use of materials that make sustainable energy supply possible.

Demonstrating that materials with a large concentration of point defects may resist fatigue due to rearrangement of the defects (stress adaptation) and/or their recombination, will allow new pathways for designing materials, experiencing periodic situations that push them significantly out of equilibrium, such as electrolytes and electrodes of fuel cells and batteries (charge-discharge) and solar energy converters (light-dark).

**Acknowledgements**:  We thank our WIS colleagues L. Kronik, O. Yaffe, D. Egger and M. Leske, as well as M. Shalom (Ben Gurion University of the Negev), and R. Lovrincic (Technical University, Braunschweig) for stimulating discussions, Ariel Biller for Fig. 3 and the Weizmann Institute of Science's Mary and Tom Beck-Canadian Centre for Alternative Energy Research for support.

# Figures

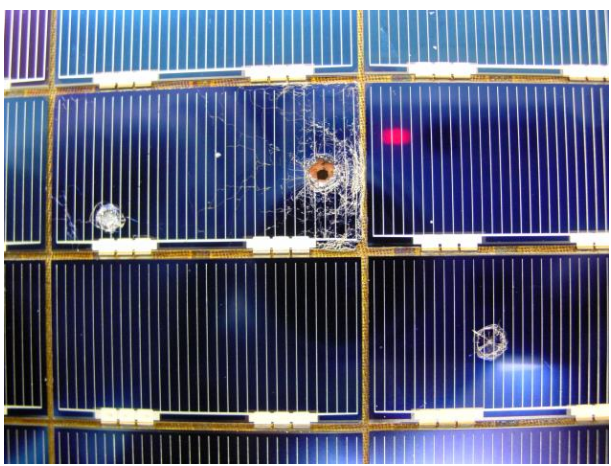

Fig. 1. ESA-built Si solar cells, retrieved from NASA's Hubble Space Telescope (lower left inset; photo: NASA) in 2002 after 8.25 years in lower earth (~ 600 km) orbit. Photos: ESA, European Space Agency.

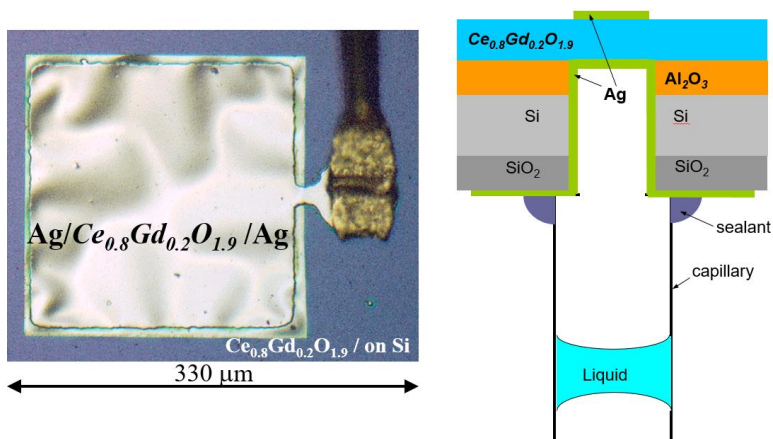

Fig. 2. Example of self-repair in anelastic material. LEFT: Top view; RIGHT: schematic cross-section of the oxygen micro-pump. Application of voltage to top contacts moves oxygen across Ce-Gd-O membrane that can be detected by movement of water droplet in the capillary. The micro-pump starts working at 600 °C but the capillary is long enough to remain at room temperature. Despite the large difference between the thermal expansion coefficients of the Si frame, CGO & Ag contacts, the micro-pump withstands hundreds of cooling-heating cycles, because below 250 °C, CGO is strongly anelastic.

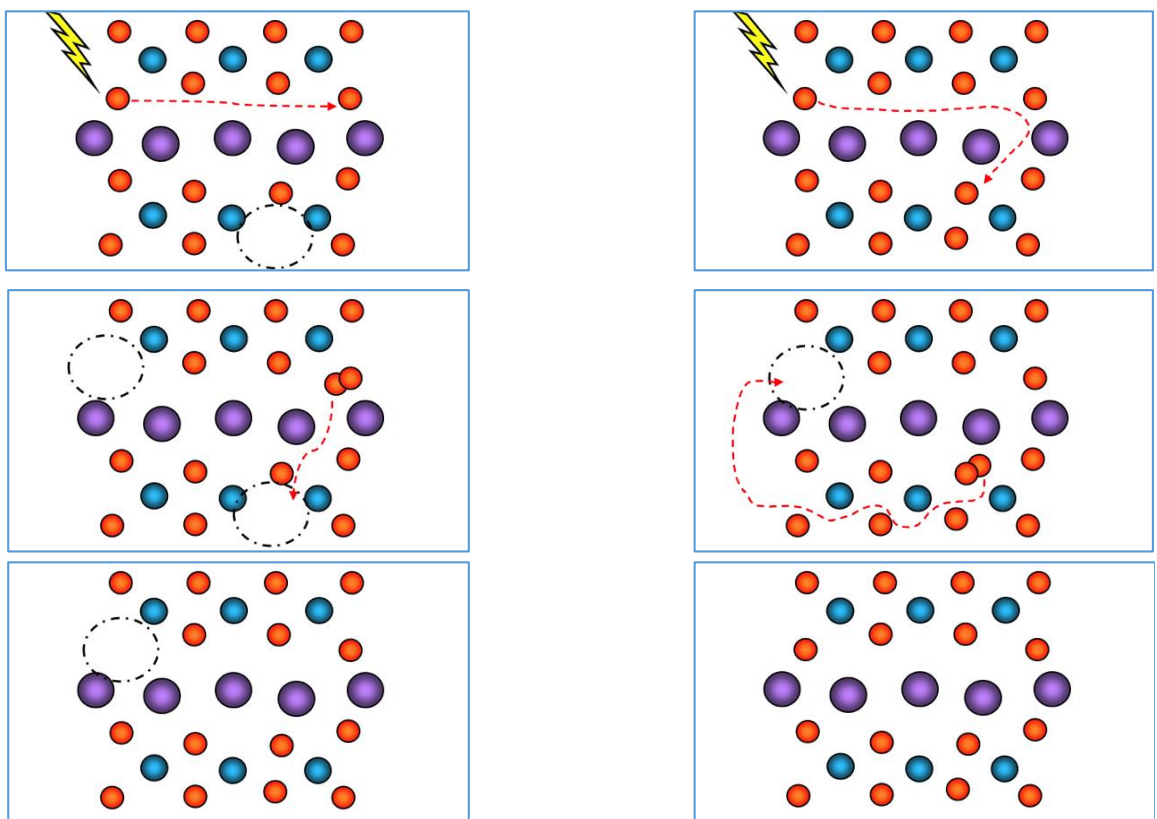

Fig. 3: Schematic illustrations (2-D projections, not to scale) of self-repair via point defect diffusion and drift, in $CuInSe_2$, the parent material for $Cu(In,Ga)Se_2$, CIGS, used in commercial thin film solar cells. Brown-orange circles: Cu; Blue circles: In; Purple circles: Se; Empty dot-dashed circle: vacancy of Cu, $V_{Cu}$, which, together with In on the Cu side, $In_{Cu}$, and Cu interstitials, $Cu_i$ are the most readily formed defects in this material; LEFT: from top to bottom: $Cu_i$ formation in material with existing vacancy and its annihilation; essentially, $V_{Cu}$ migration, but without additional defect formation; RIGHT Same but in material without vacancies, and restoration of the *status quo ante*. A slightly more realistic scheme can be found in Fig. 1 of ref. 8.